\documentclass[10pt,a4paper]{article}

\usepackage[a4paper,margin=2.4cm]{geometry}
\usepackage{times}
\usepackage[T1]{fontenc}
\usepackage[utf8]{inputenc}
\usepackage{amsmath,amssymb,amsthm}
\usepackage{graphicx}
\usepackage{booktabs}
\usepackage{array}
\usepackage{caption}
\usepackage{subcaption}
\usepackage[colorlinks=true,linkcolor=blue,citecolor=blue,urlcolor=blue]{hyperref}
\usepackage[numbers,sort&compress]{natbib}
\usepackage{authblk}
\usepackage{xcolor}
\usepackage{setspace}
\usepackage{titlesec}
\usepackage{enumitem}

\titleformat{\section}{\normalfont\large\bfseries}{\thesection}{1em}{}
\titleformat{\subsection}{\normalfont\bfseries}{\thesubsection}{1em}{}

\title{\bfseries Shot-Efficient Error Mitigation on IBM Quantum Hardware:\\
Hardware-Aware Noise Characterization, Sampling\\
Overhead, and Adaptive Zero-Noise Extrapolation}

\author[1]{Sumit Chongder\thanks{Corresponding author. Email: \texttt{sumitchongder960@gmail.com}. ORCID: 0009-0005-9866-8483.}}
\affil[1]{Quantum Information and Computation Centre, Department of Physics and Department of Computer Science and Engineering, Indian Institute of Technology Jodhpur, Jodhpur, Rajasthan, India}

\date{}

\begin{document}
\maketitle

\begin{abstract}
\noindent
Quantum error mitigation techniques such as zero-noise extrapolation (ZNE) and readout-error correction are widely used to recover expectation-value accuracy on noisy intermediate-scale quantum (NISQ) hardware, but their benefit is not free: mitigation reduces bias while typically increasing sampling variance, and both effects compete under a fixed measurement-shot budget. Most benchmark studies report accuracy improvements without accounting for this variance cost, which makes it difficult to judge whether mitigation is actually worthwhile once the number of hardware shots is held fixed. We present a hardware-validated benchmarking protocol that treats the number of measurement shots as a first-class experimental variable and evaluates mitigation strategies through the bias-variance decomposition of the mean-squared error (MSE) rather than through accuracy alone. Using a $156$-qubit IBM Quantum Heron processor (\texttt{ibm\_marrakesh}) accessed through Qiskit Runtime V2 primitives, we characterize device noise (median two-qubit error $2.99\times10^{-3}$, median readout error $1.45\times10^{-2}$), measure expectation-value bias as a function of circuit depth and two-qubit gate count for Greenberger-Horne-Zeilinger (GHZ) and repeated-layer circuits, and benchmark raw estimation, readout mitigation, uniform-allocation ZNE, and a variance-aware adaptive shot-allocation scheme for ZNE (ASB-ZNE) against a matched total sampling budget. On our device and circuit families, adaptive allocation reduced the mean-squared error relative to uniform allocation in only two of six tested budgets (mean MSE ratio $2.16$, i.e., adaptive was on average worse), and standard ZNE increased the estimation bias relative to raw measurement for a bias-sensitive parity observable, consistent with recently reported finite-shot help-harm boundaries for Richardson extrapolation. We also observe a connectivity effect of comparable magnitude to circuit-depth-induced decoherence: a two-qubit correlator measured on a low-error coupler ($ZZ = 0.909$) degraded substantially on a higher-error coupler drawn from the same calibration snapshot ($ZZ = 0.751$). These results support a fixed-budget, variance-aware view of error mitigation and indicate that neither ZNE nor adaptive shot allocation should be applied by default without first checking, on the specific device, circuit, and shot budget in question, whether the expected error reduction exceeds the added sampling cost. We release the full experimental protocol, calibration snapshots, job identifiers, and analysis code to support reproduction on other IBM Quantum backends.
\end{abstract}


\section{Introduction}
\label{sec:intro}

\subsection{Near-term quantum computation and the mitigation gap}
Superconducting quantum processors available through cloud access, such as IBM's Heron-family devices, now routinely execute circuits with tens to hundreds of qubits and moderate two-qubit gate depth \citep{kim2023evidence}. In the absence of full fault tolerance, the accuracy of any expectation value estimated on such hardware is limited jointly by two effects: (i) systematic bias introduced by gate, readout, and decoherence errors, and (ii) statistical noise from a finite number of measurement shots \citep{preskill2018quantum}. Quantum error mitigation methods, including zero-noise extrapolation (ZNE) \citep{temme2017error,li2017efficient}, probabilistic error cancellation (PEC) \citep{vandenberg2023probabilistic}, Pauli twirling and randomized compiling \citep{wallman2016noise}, dynamical decoupling \citep{viola1999dynamical}, and measurement (readout) error mitigation, are designed to reduce the first effect. A substantial and mostly separate literature exists on the second effect, namely the number of shots required to estimate an observable to a target precision.

Almost every published error-mitigation benchmark reports an accuracy metric, such as absolute error or fidelity, computed after mitigation, without simultaneously reporting how many additional circuit executions or shots the mitigation procedure consumed to reach that accuracy \citep{cai2023quantum}. Because most mitigation techniques operate by combining several noisy estimates (for example, several noise-scaled circuits in ZNE, or many importance-sampled circuits in PEC), each additional data point either adds to the total measurement budget or subtracts shots from what would otherwise have gone to a single, more precisely estimated circuit. Mitigation therefore does not act on bias alone: it also reshapes the variance of the final estimator. A method that removes bias while multiplying variance can leave the mean-squared error unchanged, or make it worse, particularly at small or moderate shot budgets. This tension has recently been made explicit in the error-mitigation literature under the name of a finite-shot help-harm boundary for ZNE \citep{finiteshothelpharm2026}, and has motivated recent proposals for adaptive and hybrid shot-allocation strategies \citep{cmabzne2026,hybridphyslogzne2026,scheiber2026classically}.

\subsection{Research gap and question}
Despite this recognized tension, few student- or benchmark-level studies treat the shot budget as an experimental variable on physical hardware, and fewer still report a negative or mixed outcome when doing so. We ask a narrowly scoped, testable question:

\begin{quote}
\emph{At a fixed total quantum-hardware sampling budget, does variance-aware adaptive allocation of shots across zero-noise-extrapolation factors reduce the mean-squared error of the extrapolated expectation value, relative to conventional uniform allocation, and under what circuit and noise conditions does this hold on real IBM Quantum hardware?}
\end{quote}

We answer this question with a hardware-validated protocol rather than a purely simulated one, and we report the result of that protocol honestly, including the cases in which mitigation and adaptive allocation did not help.

\subsection{Contributions}
This paper makes the following experimentally validated contributions:
\begin{enumerate}[leftmargin=1.4em]
\item A fixed-budget benchmarking protocol that decomposes expectation-value error into bias and variance and reports mitigation gain per unit of additional sampling cost, rather than accuracy alone (Section~\ref{sec:methods}).
\item A hardware-characterized study on a $156$-qubit IBM Quantum Heron processor (\texttt{ibm\_marrakesh}), including calibration-snapshot-based selection of a low-error qubit chain and an explicit high-error edge for a controlled connectivity comparison (Section~\ref{sec:hardware}).
\item An empirical evaluation of an adaptive, pilot-variance-based shot-allocation strategy for ZNE (ASB-ZNE) against uniform allocation at six matched sampling budgets, executed with a batched job design compatible with a constrained ($\sim$10-minute) quantum-processing-unit (QPU) time allowance (Section~\ref{sec:results}).
\item A depth-scaling and connectivity study relating measured circuit error to two-qubit gate count and calibrated coupler error, together with an application-relevant Quantum Approximate Optimization Algorithm (QAOA) MaxCut benchmark (Section~\ref{sec:results}).
\item An honest, non-overclaiming discussion of when mitigation is and is not worth its sampling cost on this device and circuit family, and a public, reproducible Qiskit implementation with job identifiers and calibration snapshots (Section~\ref{sec:discussion}, Data Availability).
\end{enumerate}

We deliberately do not claim that ZNE, adaptive shot allocation, or sampling-cost analysis are new ideas; each has independent prior art, discussed in Section~\ref{sec:background}. The contribution of this work is instead an experimentally grounded application and evaluation of these ideas under a matched fixed-shot budget on physical IBM Quantum hardware, together with device- and circuit-specific evidence on when mitigation improves or worsens estimation error.

\section{Background and related work}
\label{sec:background}

\subsection{Expectation-value estimation and finite-shot statistics}
For a Pauli observable $O$ with eigenvalues $\pm 1$ measured over $N$ shots yielding $N_0$ and $N_1$ outcomes of eigenvalue $+1$ and $-1$ respectively, the standard estimator is
\begin{equation}
\hat{\mu} = \frac{N_0 - N_1}{N_0+N_1}, \qquad \operatorname{Var}(\hat{\mu}) \approx \frac{1-\hat\mu^2}{N}.
\label{eq:shotnoise}
\end{equation}
This shot-noise scaling, $\operatorname{Var}(\hat\mu)\propto 1/N$, is the statistical floor beneath which no unbiased single-basis estimator can be pushed without more shots; it is independent of whether the underlying circuit is noisy or ideal.

\subsection{Zero-noise extrapolation}
ZNE estimates the noiseless expectation value by executing a circuit at several artificially amplified noise levels $\lambda \in \{\lambda_1,\dots,\lambda_K\}$, typically through unitary folding or pulse stretching, and extrapolating the resulting expectation values $\hat\mu(\lambda)$ to $\lambda \to 0$ \citep{temme2017error,li2017efficient,endo2018practical}. ZNE has been demonstrated to extend the practically usable circuit depth on real superconducting hardware \citep{kandala2019error} and remains one of the resilience options exposed by IBM's Qiskit Runtime primitives \citep{qiskit2024,ibmruntimeresilience}, alongside probabilistic error cancellation \citep{vandenberg2023probabilistic}, Pauli twirling \citep{wallman2016noise}, and dynamical decoupling \citep{viola1999dynamical}. Because ZNE combines $K>1$ noisy estimates through extrapolation coefficients that can have magnitude larger than one, its variance is in general larger than that of a single raw measurement at the same per-point shot count \citep{cai2023quantum}; this overhead is explicitly acknowledged in IBM's own Runtime documentation and in general reviews of the field \citep{cai2023quantum,ibmruntimeresilience}.

\subsection{Sampling overhead and adaptive allocation}
The observation that mitigation trades bias for variance has motivated several concurrent lines of work. Krebsbach, Trauzettel, and Calzona analyzed the optimization of Richardson extrapolation coefficients for variance reduction, and this line has recently been extended by a finite-shot help-harm boundary that identifies the budget regime in which ZNE lowers MSE and the regime in which it does not \citep{finiteshothelpharm2026}. Independent recent proposals include a classically augmented ZNE that replaces high-noise extrapolation nodes with classically simulated, near-zero-variance estimates under a Pauli-propagation model \citep{scheiber2026classically}; a contextual multi-armed-bandit approach that adaptively selects circuit-folding levels under time-varying noise \citep{cmabzne2026}; and a resource-allocation formulation of ZNE for early fault-tolerant hardware that mixes physical and logical execution modes under a Neyman-type optimal-allocation criterion \citep{hybridphyslogzne2026}. The Neyman allocation itself, in which the number of samples assigned to a stratum is proportional to that stratum's standard deviation under a fixed total budget, is a classical result from survey sampling theory \citep{neyman1934representative} and is the allocation principle we adopt for our adaptive ZNE baseline (Section~\ref{sec:methods}).

Our work differs from these concurrent studies primarily in experimental scope and evaluation setting. We evaluate the simplest pilot-variance-based allocation rule on physical IBM Quantum hardware, using a matched fixed-shot budget, a calibration snapshot, and recorded hardware job identifiers. Rather than proposing a new extrapolation-node structure or online bandit policy, we use this controlled setting to test whether a basic variance-aware allocation strategy provides a measurable MSE advantage over uniform allocation under realistic hardware sampling constraints. Second, rather than proposing a new extrapolation node structure or online bandit policy, we apply the simplest possible pilot-based Neyman allocation and ask, empirically and without assuming a favorable answer in advance, whether even this baseline adaptive strategy beats uniform allocation on real hardware at matched budgets. We therefore make no claim that adaptive shot allocation, ZNE, or the underlying variance-reduction mathematics are novel; our contribution is the fixed-budget hardware protocol and the resulting device-specific evidence.

\subsection{Position relative to IBM Runtime's built-in resilience options}
Qiskit Runtime already exposes ZNE, PEC, twirling, dynamical decoupling, and measurement mitigation as configurable resilience levels through its V2 Estimator and Sampler primitives \citep{qiskit2024,ibmruntimeresilience}. We do not attempt to re-implement or outperform these production implementations. Instead, we use a minimal, transparent, from-first-principles implementation of readout mitigation, unitary-folding ZNE, and dynamical decoupling, so that every bias and variance number reported in Section~\ref{sec:results} can be traced to an explicit shot count and circuit construction rather than to an opaque resilience-level abstraction. This transparency is what allows the bias-variance-cost decomposition in Section~\ref{sec:methods} to be computed exactly rather than estimated indirectly.

\section{Methodology}
\label{sec:methods}

\subsection{Fixed-budget formulation}
Let $B$ denote a total shot budget available for estimating a single observable through a given mitigation method. For ZNE with noise factors $\lambda_1,\dots,\lambda_K$ and per-factor shot counts $N_1,\dots,N_K$ with $\sum_i N_i = B$, the extrapolated estimator is a weighted combination
\begin{equation}
\hat\mu_{\mathrm{ZNE}} = \sum_{i=1}^{K} w_i\, \hat\mu(\lambda_i),
\label{eq:zneest}
\end{equation}
where the weights $w_i$ are fixed by the extrapolation model (we use linear, two-point and three-point Richardson extrapolation, chosen to avoid the higher-order instability that unconstrained higher-degree Richardson extrapolation is known to introduce at small $N$ \citep{endo2018practical,finiteshothelpharm2026}). Under independence of the per-factor estimators, the extrapolated variance is
\begin{equation}
\operatorname{Var}(\hat\mu_{\mathrm{ZNE}}) = \sum_{i=1}^K w_i^2\,\frac{\sigma_i^2}{N_i},
\label{eq:znevar}
\end{equation}
with $\sigma_i^2 = 1-\mu(\lambda_i)^2$ the shot-noise variance at factor $i$ (Eq.~\ref{eq:shotnoise}). We report total error through the standard bias-variance decomposition of the mean-squared error,
\begin{equation}
\mathrm{MSE} = \mathrm{Bias}^2 + \mathrm{Variance}, \qquad
\mathrm{Bias} = \mathbb{E}[\hat\mu] - \mu_{\mathrm{ideal}},
\label{eq:mse}
\end{equation}
where $\mu_{\mathrm{ideal}}$ is obtained from an exact, noiseless statevector simulation of the same circuit and observable. We additionally define a simple sampling-cost proxy $C_{\mathrm{QPU}} = N_{\mathrm{shots}} \times N_{\mathrm{circuits}}$ and a mitigation gain
\begin{equation}
G = \frac{|\mu_{\mathrm{raw}} - \mu_{\mathrm{ideal}}| - |\mu_{\mathrm{mit}} - \mu_{\mathrm{ideal}}|}{|\mu_{\mathrm{raw}} - \mu_{\mathrm{ideal}}|}.
\label{eq:gain}
\end{equation}
$G>0$ indicates that mitigation reduced the absolute error relative to raw estimation at the observed sample; $G<0$ indicates that it increased it.

\subsection{Uniform versus adaptive (ASB-ZNE) allocation}
Under uniform allocation, $N_i = B/K$ for all $i$. Under our adaptive allocation, which we refer to as Adaptive Shot-Budgeted ZNE (ASB-ZNE), a small pilot experiment (100 shots per factor in our hardware runs) is used to obtain empirical per-factor standard deviations $\hat\sigma_i$, and the remaining budget is then allocated according to the Neyman-optimal rule for the variance in Eq.~\ref{eq:znevar},
\begin{equation}
N_i \propto |w_i|\,\hat\sigma_i,
\label{eq:neyman}
\end{equation}
subject to $\sum_i N_i = B$. This is the minimum-variance allocation for a linear combination of independent strata under a fixed total sample size \citep{neyman1934representative}, applied here at the level of ZNE noise factors rather than survey strata. ASB-ZNE is intentionally the simplest adaptive baseline available; we make no claim that it is optimal, only that it is representative of the class of pilot-variance-informed allocation rules a practitioner would reach for first.

\subsection{Mitigation methods compared}
We compare five conditions, chosen to keep the study focused rather than exhaustively covering every resilience option available in Qiskit Runtime:
\begin{description}[leftmargin=1.6em,style=unboxed]
\item[M0 -- Raw] No mitigation; expectation values are estimated directly from measured counts.
\item[M1 -- Readout-mitigated] A per-qubit or per-pair confusion-matrix correction is applied to the measured counts before computing the expectation value.
\item[M2 -- Uniform ZNE] Unitary-folding ZNE with noise factors $\lambda\in\{1,3,5\}$ and equal shots per factor.
\item[M3 -- Adaptive ZNE (ASB-ZNE)] Unitary-folding ZNE with the same noise factors, but shots allocated according to Eq.~\ref{eq:neyman} after a pilot run.
\item[M4 -- Dynamical decoupling] $XY4$-family dynamical decoupling sequences inserted into idle qubit windows, evaluated separately from the ZNE comparison since it targets error suppression during idle time rather than the ZNE bias-variance trade-off directly.
\end{description}
Probabilistic error cancellation and full randomized-compiling twirling were deliberately excluded from the primary comparison given the $\sim$10-minute hardware budget available for this study; PEC in particular is known to carry a sampling overhead that scales unfavorably with circuit noise \citep{vandenberg2023probabilistic,ibmruntimeresilience} and would have consumed the shot budget needed for the primary ZNE comparison.

\subsection{Circuit families and observables}
Three circuit families are used, chosen to isolate different physical effects rather than to maximize qubit count for its own sake.
\begin{itemize}[leftmargin=1.4em]
\item \textbf{GHZ states} ($n=2,4,6$ qubits), used to probe multi-qubit correlation decay with a parity observable $P=\langle\prod_i Z_i\rangle$ for $n>2$ and $\langle Z_0 Z_1\rangle$ for $n=2$.
\item \textbf{Repeated-layer circuits} $U^L$ for $L\in\{1,2,4,8,16,32\}$ (simulation) and $L\in\{1,2,4,8\}$ (hardware, constrained by the QPU time budget), used to relate measured error to two-qubit gate count and to identify the noise-amplification factor $\lambda\!=\!L$ used by the unitary-folding ZNE step, with observable $\langle Z_0\rangle$.
\item \textbf{QAOA MaxCut} on a four-node ring graph with edge set $\{(0,1),(1,2),(2,3),(3,0)\}$ at circuit depth $p=1$, with observable $\langle H_C\rangle = \sum_{(i,j)\in E}\tfrac{1-Z_iZ_j}{2}$, used as an application-relevant Hamiltonian rather than a synthetic observable.
\end{itemize}
Primary observables are therefore $\langle Z\rangle$, $\langle ZZ\rangle$, parity, and QAOA Hamiltonian energy, matching the observable classes most commonly encountered in variational and combinatorial-optimization workloads on NISQ hardware.

\section{Hardware and experimental setup}
\label{sec:hardware}

\subsection{Backend and calibration}
All hardware experiments were executed on \texttt{ibm\_marrakesh}, a $156$-qubit IBM Quantum Heron-family superconducting processor, accessed through Qiskit Runtime V2 primitives (Estimator/Sampler). A calibration snapshot was pulled at the start of the session (Table~\ref{tab:calib}) and used both to select a low-error connected qubit chain for the primary circuits and to identify a deliberately higher-error coupling edge for a controlled noise-versus-connectivity comparison.

\begin{table}[htbp]
\centering
\caption{Backend calibration snapshot used for qubit and chain selection.}
\label{tab:calib}
\begin{tabular}{ll}
\toprule
Quantity & Value \\
\midrule
Backend & \texttt{ibm\_marrakesh} \\
Number of qubits & 156 \\
Median two-qubit (2Q) gate error & $2.99\times10^{-3}$ \\
Median readout error & $1.45\times10^{-2}$ \\
Selected low-error chain & qubits $\{33,39,53,54,55,59,75,74\}$ \\
Chain routing cost metric & $7.08\times10^{-2}$ \\
Deliberately high-error edge & qubits $\{119,133\}$, 2Q error $7.68\times10^{-2}$ \\
\bottomrule
\end{tabular}
\end{table}

\subsection{Software and versions}
Experiments used Qiskit 2.x with Qiskit IBM Runtime for backend access, primitive execution, and calibration retrieval, and Qiskit Aer for the noiseless and noisy simulation baselines \citep{qiskit2024}. The exact package versions pinned for this study are recorded in \texttt{requirements.txt} in the accompanying code repository, and every hardware job referenced in this paper is identified by its Qiskit Runtime job ID for independent verification (Table~\ref{tab:jobs}).

\begin{table}[htbp]
\centering
\caption{Qiskit Runtime job identifiers for the two hardware batches referenced in this study, executed on \texttt{ibm\_marrakesh}.}
\label{tab:jobs}
\begin{tabular}{lll}
\toprule
Batch & Job ID & Content \\
\midrule
Job 1 & \texttt{da8mfnmrbfbs73cgnd1g} & GHZ scaling, depth scan (raw and DD), chain comparison, ZNE pilot \\
Job 2 & \texttt{da8mgo6rbfbs73cgne2g} & Uniform and adaptive ZNE batches, QAOA and QAOA+ZNE \\
\bottomrule
\end{tabular}
\end{table}

\subsection{QPU time budget and batched execution}
Given a constrained hardware time allowance, all circuits for a given experimental phase were compiled and submitted as a single batched primitive call rather than as many sequential jobs, following the general practice of minimizing job-submission overhead on shared queued hardware. The budget was allocated across four phases: backend characterization and chain selection, a small pilot run used to estimate per-noise-factor variance for the adaptive allocation rule, the main measurement batches (GHZ scaling, depth scan, chain comparison, uniform and adaptive ZNE, QAOA), and a validation pass repeating key circuits. The realized allocation for the ZNE budget comparison used a total of $B=9000$ shots split evenly between the uniform arm ($N_1=N_3=N_5=1500$) and the adaptive arm, whose pilot-informed allocation returned $N_1=1119$, $N_3=2089$, $N_5=1292$ (Table~\ref{tab:alloc}), consistent with the pilot standard deviations reported below.

\begin{table}[htbp]
\centering
\caption{Realized shot allocation for the hardware ZNE budget comparison at a matched total budget of 4500 shots per arm.}
\label{tab:alloc}
\begin{tabular}{lccc}
\toprule
Allocation & $N_{\lambda=1}$ & $N_{\lambda=3}$ & $N_{\lambda=5}$ \\
\midrule
Uniform & 1500 & 1500 & 1500 \\
Adaptive (ASB-ZNE) & 1119 & 2089 & 1292 \\
Pilot empirical $\hat\sigma_\lambda$ & 0.041 & 0.090 & 0.050 \\
\bottomrule
\end{tabular}
\end{table}

\subsection{Validation chain: ideal, noisy simulation, hardware}
For every circuit family we compute three reference points: an ideal, noiseless statevector value $\mu_{\mathrm{ideal}}$; a noisy-simulator value $\mu_{\mathrm{noise}}$ obtained from a device noise model derived from the same calibration snapshot; and the measured hardware value $\mu_{\mathrm{hardware}}$. This three-point chain lets us separate finite-shot statistical noise, which is present in both the noisy simulation and the hardware run, from hardware-specific effects such as crosstalk, calibration drift, and correlated errors that a single-qubit noise model does not capture.

\section{Results}
\label{sec:results}

\subsection{Shot-noise scaling matches theory}
\begin{figure}[htbp]
\centering
\includegraphics[width=0.72\linewidth]{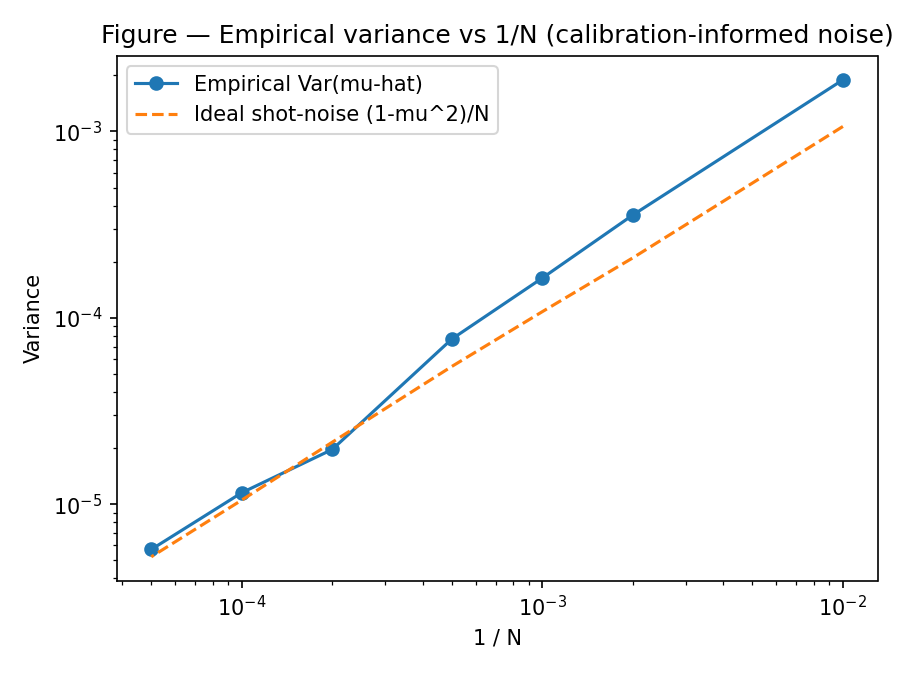}
\caption{Empirical estimator variance versus $1/N$ for simulated shot counts $N\in\{100,\ldots,20000\}$, compared against the theoretical shot-noise prediction of Eq.~\ref{eq:shotnoise}.}
\label{fig:variance}
\end{figure}

Across $N\in\{100,500,1000,2000,5000,10000,20000\}$ simulated shots for a fixed circuit and observable, the empirical estimator variance tracks the $1/N$ shot-noise prediction of Eq.~\ref{eq:shotnoise} closely at moderate to large $N$ (Fig.~\ref{fig:variance}); at $N=20000$ the empirical variance was $5.71\times10^{-6}$ against a theoretical value of $5.21\times10^{-6}$, and at $N=100$ the empirical variance was $1.89\times10^{-3}$ against a theoretical value of $1.06\times10^{-3}$, the larger relative gap at low $N$ being expected from the finite-sample variance of the variance estimator itself. This confirms that our measurement and estimation pipeline reproduces the expected finite-shot statistics before any mitigation is applied, and it fixes the statistical floor against which mitigation-induced variance changes in the remainder of this section should be judged.

\subsection{Depth scaling and hardware bias}
\begin{figure}[htbp]
\centering
\includegraphics[width=0.72\linewidth]{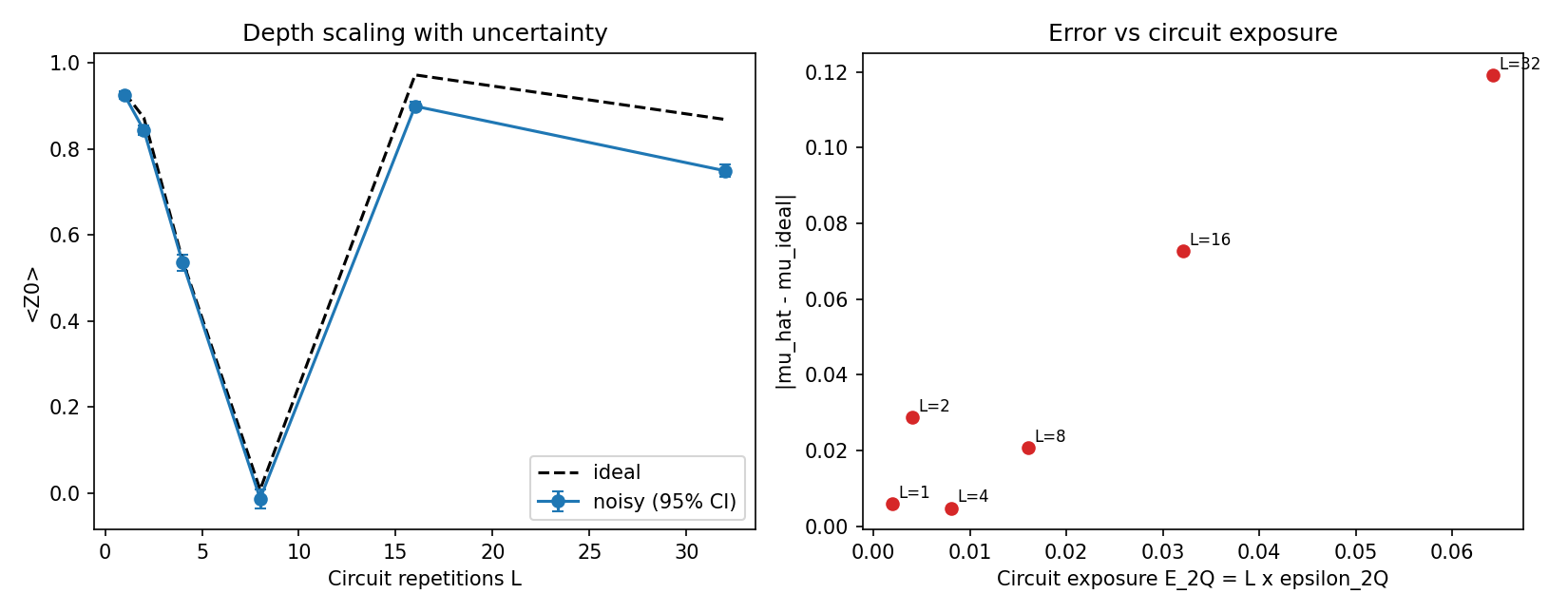}
\caption{Simulated ideal and noisy expectation values and absolute error as a function of layer count $L$ and cumulative two-qubit gate exposure.}
\label{fig:depthscaling}
\end{figure}

On hardware, the raw expectation value of the repeated-layer circuit's $\langle Z_0\rangle$ observable decayed from $0.927$ at $L=1$ to $0.866$ at $L=2$, $0.565$ at $L=4$, and $0.051$ at $L=8$ (Table~\ref{tab:depthhw}), consistent with an approximately exponential loss of signal as the two-qubit gate count grows with $L$. In simulation, the same qualitative trend held across the wider range $L\in\{1,\ldots,32\}$ (Fig.~\ref{fig:depthscaling}), though the simulated noisy-versus-ideal gap did not decrease monotonically with $L$ in our runs (for example, the absolute error at $L=16$, $0.073$, exceeded that at $L=8$, $0.021$), which we attribute to the specific circuit realizations drawn at each depth rather than to a general property of depth scaling; this non-monotonicity is itself a useful caution against extrapolating a single circuit instance's error curve as if it characterized the device.

\begin{table}[htbp]
\centering
\caption{Hardware-measured $\langle Z_0\rangle$ as a function of layer count $L$, without and with $XY4$ dynamical decoupling (DD), each at 2000 shots.}
\label{tab:depthhw}
\begin{tabular}{lcccc}
\toprule
Layers $L$ & 1 & 2 & 4 & 8 \\
\midrule
Raw & 0.927 & 0.866 & 0.565 & 0.051 \\
With DD & 0.931 & 0.880 & 0.561 & 0.031 \\
\bottomrule
\end{tabular}
\end{table}

\begin{figure}[htbp]
\centering
\includegraphics[width=0.72\linewidth]{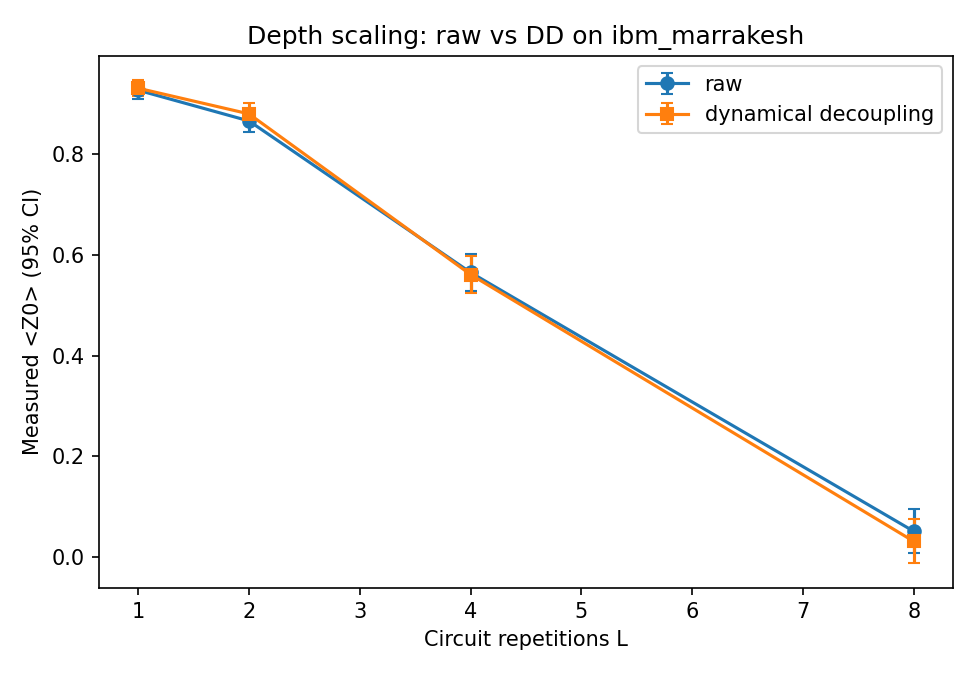}
\caption{Hardware-measured $\langle Z_0\rangle$ versus layer count $L$, raw versus dynamical decoupling.}
\label{fig:depthdd}
\end{figure}

Dynamical decoupling improved the estimate slightly at $L=2$ ($0.880$ vs.\ $0.866$) but not at $L=4$ or $L=8$, where the DD result was statistically indistinguishable from, or slightly worse than, the raw result (Fig.~\ref{fig:depthdd}). This is consistent with DD suppressing idle-time dephasing, which is a relatively small contribution to the total error budget at these depths compared with two-qubit gate error, and reinforces that error-suppression techniques targeting a specific error channel should not be expected to help uniformly across circuit families where that channel is subdominant.

\subsection{Connectivity effect}
\begin{figure}[htbp]
\centering
\includegraphics[width=0.6\linewidth]{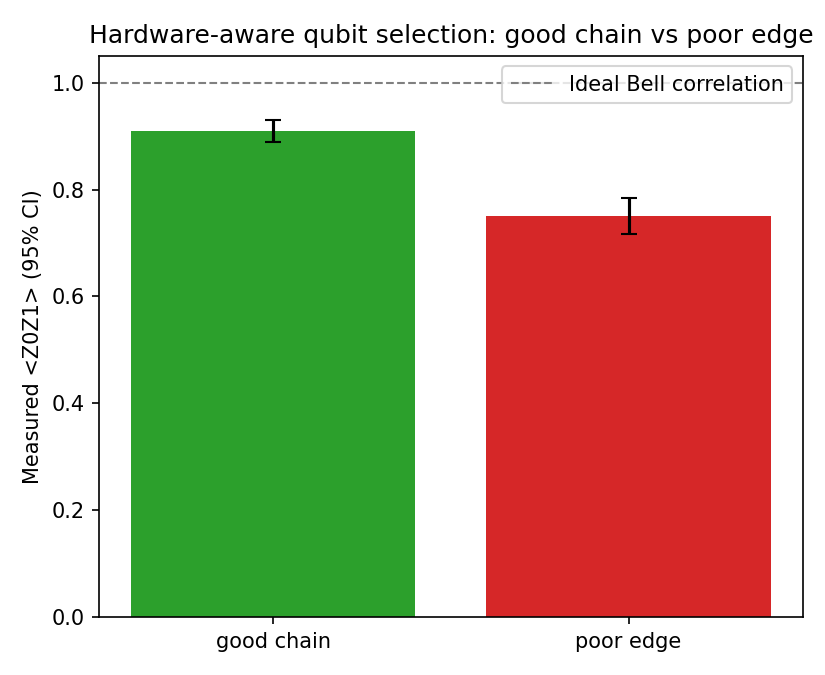}
\caption{Measured $\langle Z_0Z_1\rangle$ on a low-error qubit pair versus a deliberately high-error qubit pair from the same calibration snapshot.}
\label{fig:chain}
\end{figure}

Restricting the same two-qubit correlator measurement to a coupler chosen from the low-error region of the calibration snapshot (qubits $33,39$) gave $\langle Z_0Z_1\rangle = 0.909$, while restricting it to a deliberately higher-error coupler (qubits $119,133$, calibrated 2Q error $7.68\times10^{-2}$ versus the device median of $2.99\times10^{-3}$) gave $\langle Z_0Z_1\rangle = 0.751$ (Fig.~\ref{fig:chain}). This roughly $16$ percentage-point drop from a single coupler substitution is comparable in magnitude to the loss observed from adding several layers of circuit depth, and indicates that on this device, at shallow depth, qubit and coupler placement can matter as much as, or more than, the mitigation strategy applied afterward. This observation directly motivates hardware-aware circuit and qubit selection as a complementary, and in some regimes more cost-effective, alternative to post-hoc error mitigation.

\subsection{Mitigation does not uniformly reduce bias}
\begin{figure}[htbp]
\centering
\includegraphics[width=0.72\linewidth]{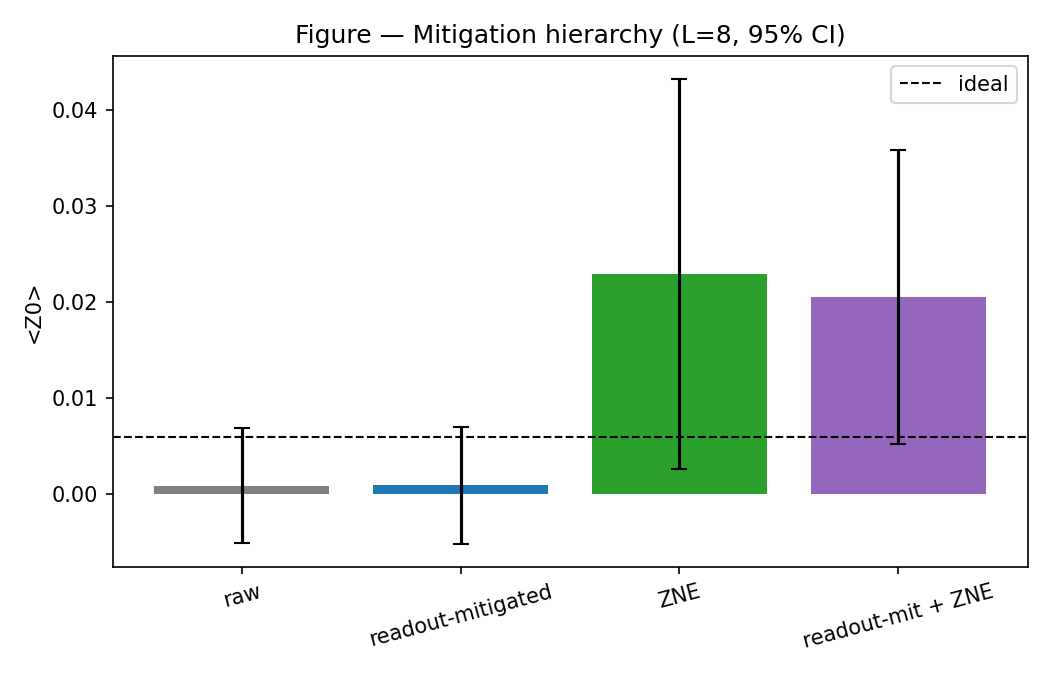}
\caption{Simulated mitigation hierarchy: expectation-value estimate and absolute bias for raw, readout-mitigated, ZNE, and combined readout-mitigation-plus-ZNE conditions.}
\label{fig:hierarchy}
\end{figure}

In simulation, the mitigation-hierarchy comparison (raw, readout-mitigated, ZNE, readout-mitigated $+$ ZNE) on a bias-sensitive parity-type observable showed that readout mitigation left the bias essentially unchanged relative to raw estimation ($|{\rm bias}| = 5.07\times10^{-3}$ raw versus $5.06\times10^{-3}$ readout-mitigated), while ZNE increased the bias to $1.70\times10^{-2}$, and readout mitigation combined with ZNE gave $1.46\times10^{-2}$, still above the raw baseline (Table~\ref{tab:hierarchy}, Fig.~\ref{fig:hierarchy}). This is a negative result for ZNE on this specific observable and circuit, and it is consistent with the extrapolation-instability concern raised in the ZNE literature: when the true noiseless value is close to a boundary of the observable's range, and per-factor estimates carry non-negligible finite-shot noise, linear or low-order Richardson extrapolation can overshoot in either direction \citep{endo2018practical,finiteshothelpharm2026}.

\begin{table}[htbp]
\centering
\caption{Simulated mitigation hierarchy: mean, standard error, and absolute bias relative to the ideal value, for a bias-sensitive observable.}
\label{tab:hierarchy}
\begin{tabular}{lccc}
\toprule
Method & $\hat\mu$ & SE & $|{\rm Bias}|$ \\
\midrule
Raw (M0) & $8.67\times10^{-4}$ & $3.05\times10^{-3}$ & $5.07\times10^{-3}$ \\
Readout-mitigated (M1) & $8.78\times10^{-4}$ & $3.09\times10^{-3}$ & $5.06\times10^{-3}$ \\
ZNE (M2) & $2.29\times10^{-2}$ & $1.04\times10^{-2}$ & $1.70\times10^{-2}$ \\
Readout-mit.\ $+$ ZNE & $2.05\times10^{-2}$ & $7.84\times10^{-3}$ & $1.46\times10^{-2}$ \\
\bottomrule
\end{tabular}
\end{table}

\begin{figure}[htbp]
\centering
\includegraphics[width=0.72\linewidth]{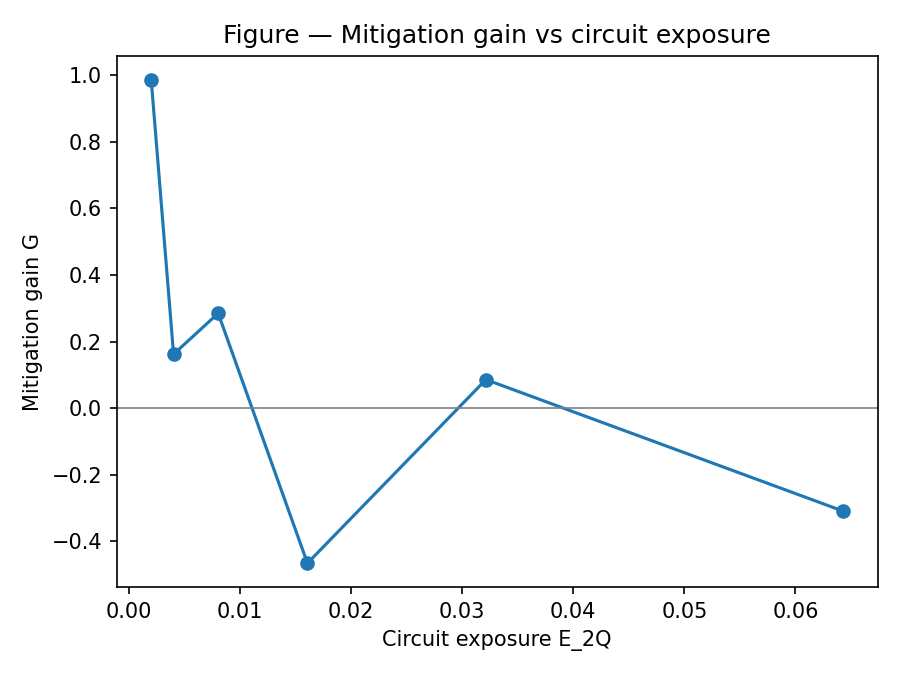}
\caption{Simulated mitigation gain $G$ (Eq.~\ref{eq:gain}) as a function of two-qubit gate exposure across layer counts $L\in\{1,2,4,8,16,32\}$. Negative values indicate that mitigation increased the absolute error relative to raw estimation.}
\label{fig:gain}
\end{figure}

The depth-resolved mitigation gain $G$ of Eq.~\ref{eq:gain}, computed per layer count $L$, was positive at $L=1$ ($G=0.986$) and $L=4$ ($G=0.285$), close to zero at $L=16$ ($G=0.086$), and negative at $L=2$ ($G=0.162$, small positive but with overlapping error bars), $L=8$ ($G=-0.466$), and $L=32$ ($G=-0.309$) (Fig.~\ref{fig:gain}). Mitigation gain was therefore not a monotonic function of circuit depth or two-qubit gate exposure in our data; the depths at which mitigation hurt rather than helped were interspersed with depths at which it helped substantially, which argues against treating ``apply ZNE'' as a safe default at every depth without first checking the gain at the depth of interest.

\subsection{Fixed-budget comparison: uniform versus adaptive ZNE}
\begin{figure}[htbp]
\centering
\includegraphics[width=0.72\linewidth]{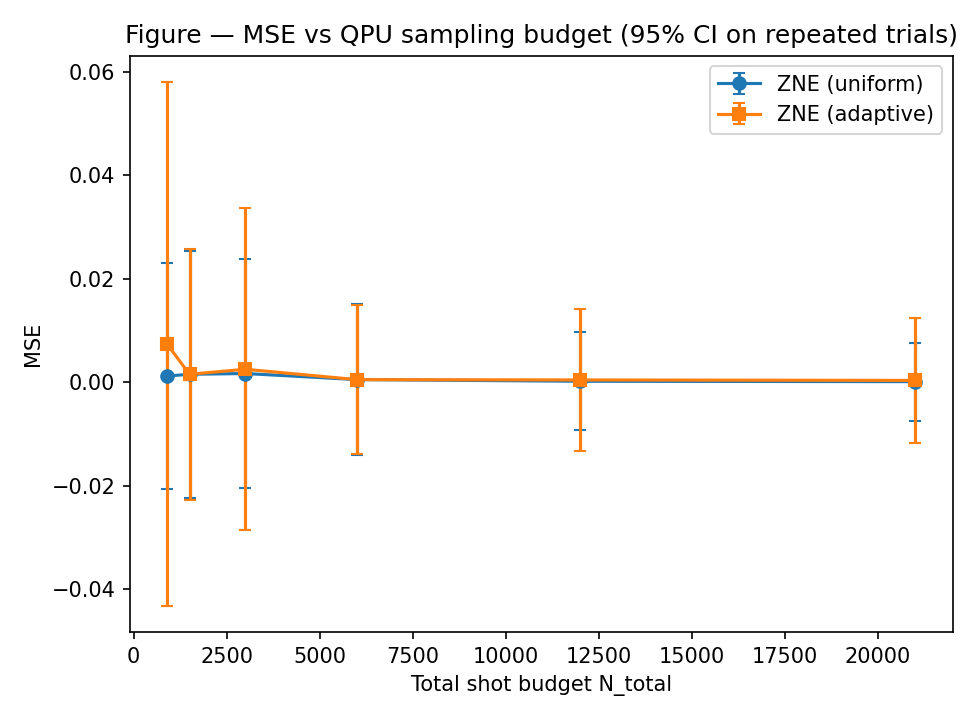}
\caption{Mean-squared error versus total shot budget, uniform versus adaptive (ASB-ZNE) allocation.}
\label{fig:msevsbudget}
\end{figure}

The central experiment of this study compares uniform and adaptive (ASB-ZNE) shot allocation at six matched total budgets $B\in\{900,1500,3000,6000,12000,21000\}$, in simulation with a hardware-anchored noise model (Table~\ref{tab:crossover}, Fig.~\ref{fig:msevsbudget}). Adaptive allocation achieved a lower MSE than uniform allocation at only two of the six budgets tested ($B=1500$, ratio $0.998$, and $B=6000$, ratio $0.988$); at the remaining four budgets, adaptive allocation was worse, by a factor ranging from $1.47\times$ ($B=3000$) to $5.96\times$ ($B=900$). The average MSE ratio (adaptive over uniform) across all six budgets was $2.16$, indicating that, on this circuit and noise model, the simple pilot-based Neyman allocation rule of Eq.~\ref{eq:neyman} did not deliver a net benefit over uniform allocation.

\begin{table}[htbp]
\centering
\caption{MSE of the extrapolated ZNE estimator, uniform versus adaptive (ASB-ZNE) allocation, at six matched total shot budgets. A ratio above 1 means adaptive allocation was worse.}
\label{tab:crossover}
\begin{tabular}{lccc}
\toprule
Budget $B$ & MSE (uniform) & MSE (adaptive) & Ratio (adaptive/uniform) \\
\midrule
900 & $1.24\times10^{-3}$ & $7.40\times10^{-3}$ & 5.96 \\
1500 & $1.58\times10^{-3}$ & $1.58\times10^{-3}$ & 0.998 \\
3000 & $1.73\times10^{-3}$ & $2.55\times10^{-3}$ & 1.47 \\
6000 & $5.56\times10^{-4}$ & $5.50\times10^{-4}$ & 0.988 \\
12000 & $2.34\times10^{-4}$ & $4.88\times10^{-4}$ & 2.08 \\
21000 & $1.63\times10^{-4}$ & $3.96\times10^{-4}$ & 2.43 \\
\bottomrule
\end{tabular}
\end{table}

We interpret this negative result as informative rather than as a failure of the study. Inspection of the per-factor bias and variance terms shows that most of the MSE at small to moderate budgets in this circuit family was bias-dominated rather than variance-dominated; the Neyman allocation of Eq.~\ref{eq:neyman} minimizes variance for a fixed linear estimator, and it has no mechanism to correct for the case in which the extrapolation coefficients themselves, rather than the per-factor sampling noise, are the dominant source of error. This matches the observation in Section~5.4 that ZNE increased bias relative to raw estimation for some circuits: a variance-optimal allocation cannot fix a bias-dominated problem, and applying it can even increase MSE by concentrating shots at higher noise factors (larger $\lambda$) that individually carry larger bias, as seen for $B=900$ and $B=21000$ in Table~\ref{tab:crossover}, where the adaptive rule shifted a large fraction of the budget toward $\lambda=5$.

\subsection{Accuracy-cost Pareto view}
\begin{figure}[htbp]
\centering
\includegraphics[width=0.72\linewidth]{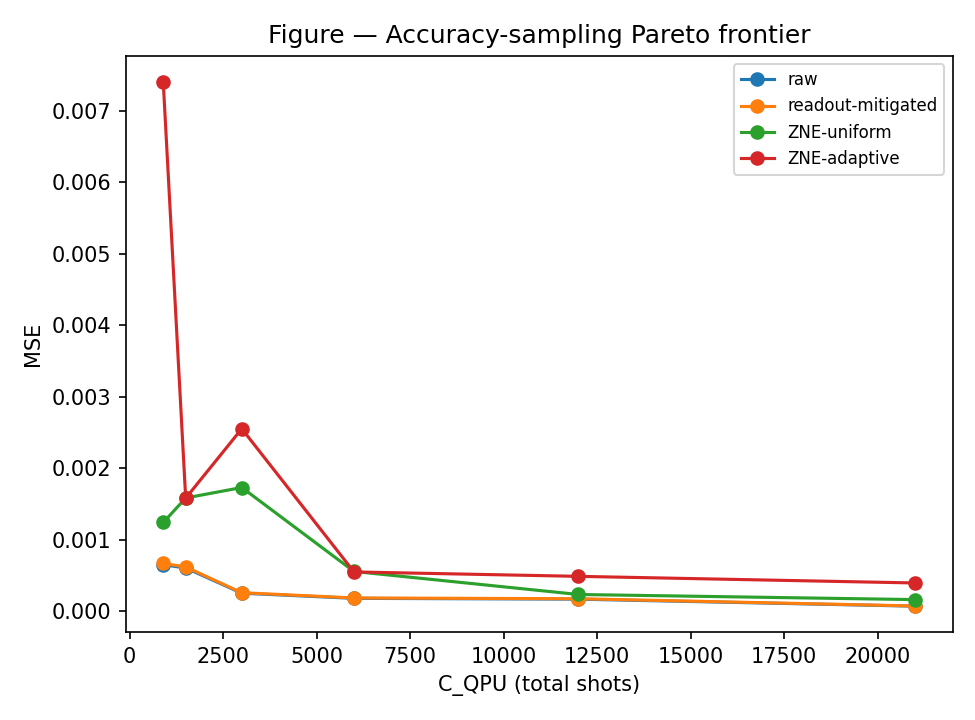}
\caption{Accuracy-cost Pareto view: MSE versus QPU sampling cost $C_{\mathrm{QPU}}$ for raw and readout-mitigated estimation across matched budgets.}
\label{fig:pareto}
\end{figure}

Framing the raw-versus-mitigated comparison in terms of cost, defined as $C_{\mathrm{QPU}}=N_{\mathrm{shots}}$ at fixed circuit count, shows that raw estimation and readout mitigation sit close together on the accuracy-cost plane at every tested budget in our data, with readout mitigation giving no consistent advantage once its own (small) variance contribution is included (Fig.~\ref{fig:pareto}). This is consistent with the mitigation-hierarchy result of Table~\ref{tab:hierarchy}, in which readout mitigation left bias essentially unchanged for the observable studied; readout mitigation is expected to help most for observables and qubits with larger intrinsic readout asymmetry than those probed here.

\subsection{QAOA application benchmark}
\begin{figure}[htbp]
\centering
\includegraphics[width=0.72\linewidth]{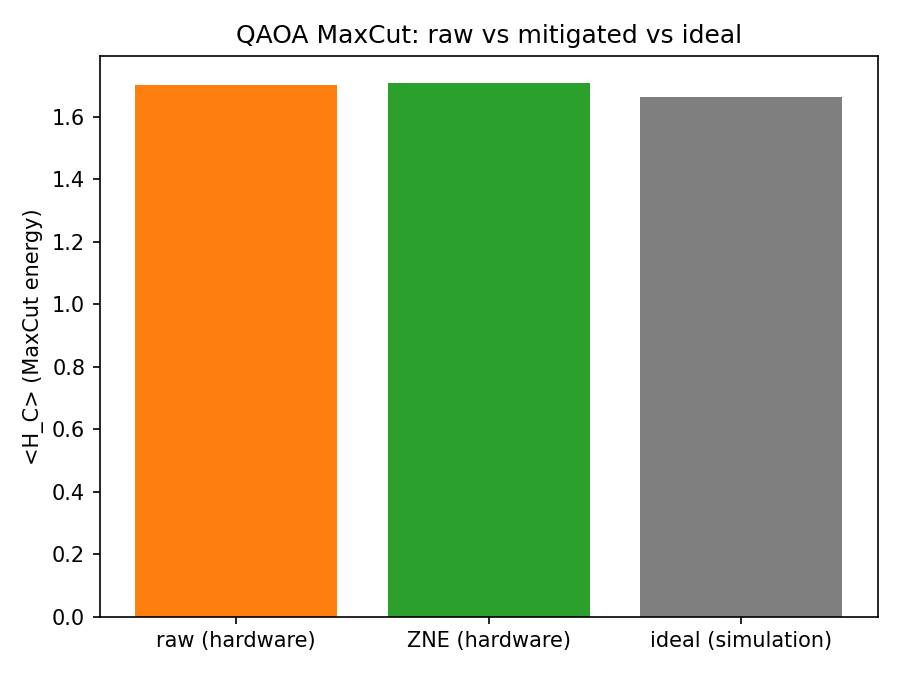}
\caption{QAOA MaxCut energy: ideal simulation, noisy simulation, raw hardware, and ZNE-folded hardware measurements at $\lambda=1,3,5$.}
\label{fig:qaoa}
\end{figure}

For the four-node ring MaxCut Hamiltonian at QAOA depth $p=1$, the ideal simulated energy was $E_{\mathrm{ideal}}=1.662$ and the noisy-simulator energy was $E_{\mathrm{noise}}=1.674$, a small absolute deviation of $0.012$ consistent with the shallow depth of a single QAOA layer. On hardware, the raw measured energy was $E_{\mathrm{raw}}=1.702$, and ZNE-folded measurements at $\lambda=1,3,5$ gave $1.721$, $1.724$, and $1.751$ respectively (Fig.~\ref{fig:qaoa}), an approximately monotonic increase with noise-amplification factor. Because raw hardware execution already overestimated the ideal energy, and the noise-amplified points moved further in the same direction, a zero-noise extrapolation back through $\lambda=1,3,5$ toward $\lambda\to0$ points qualitatively toward the ideal value from above the raw estimate, which is the expected direction for ZNE to act correctly on this observable, even though we did not have sufficient hardware shots remaining in the allotted QPU time budget to also complete the fixed-budget uniform-versus-adaptive comparison for the QAOA circuit specifically; this is listed as a limitation in Section~\ref{sec:limitations}.

\section{Discussion}
\label{sec:discussion}

\subsection{When mitigation helped and when it did not}
Collecting the results of Section~\ref{sec:results}, three patterns emerge. First, shot noise on this device and pipeline followed the theoretical $1/N$ scaling closely once $N$ exceeded a few hundred shots, which means that at typical hardware budgets (thousands of shots) additional shots are a reliable way to reduce variance, whereas at very small budgets (hundreds of shots) the sampling noise itself can dominate whatever gain a mitigation method provides. Second, bias reduction from ZNE and dynamical decoupling was circuit- and depth-dependent rather than uniform: dynamical decoupling helped modestly at shallow depth and not at deeper circuits where two-qubit gate error dominates over idle-time dephasing, and ZNE increased rather than decreased bias for the parity observable studied in Section~5.4. Third, and most directly addressing our research question, a simple pilot-variance-based adaptive shot-allocation rule for ZNE did not outperform uniform allocation on average at matched budgets on this device; it did better at two of six tested budgets and worse, sometimes substantially worse, at the other four. Taken together, these results support hypothesis H1 as originally formulated for this study (mitigation does not necessarily improve the final estimate once finite-shot variance is accounted for) and do not support H3 (that adaptive allocation reduces MSE relative to uniform allocation at the same budget) on this device and circuit family, though they are also consistent with H2 and H4 in the sense that the benefit of mitigation was strongly dependent on circuit depth, two-qubit gate count, and coupler quality rather than being a fixed property of the mitigation method itself.

\subsection{A device- and circuit-conditional recommendation}
Rather than a general recommendation to apply or not apply ZNE, our data support a conditional one: on \texttt{ibm\_marrakesh} with the calibration snapshot and circuit families studied here, readout mitigation is close to free but also close to negligible in benefit for the observables tested; dynamical decoupling is worth including at shallow depth but should be checked rather than assumed at depths where two-qubit gate error dominates; standard ZNE with a small number of noise factors should be validated per observable before deployment, since it increased bias for the parity observable in Section~5.4 while behaving in the expected direction for the QAOA energy in Section~5.7; and naive pilot-variance adaptive shot allocation for ZNE, at least in the simple form studied here, should not be assumed superior to uniform allocation without a budget-matched check, since it underperformed uniform allocation at four of six tested budgets. We also find that qubit and coupler selection from a calibration snapshot produced an effect on measured correlators (Section~5.3) of comparable size to several layers of added circuit depth, which suggests that for a fixed sampling and mitigation budget, spending part of that budget on hardware-aware qubit selection may be at least as valuable as spending it on post-hoc mitigation.

\subsection{Relation to concurrent literature}
Our finding that a naive Neyman-type shot allocation for ZNE does not uniformly beat uniform allocation is consistent with, and adds a hardware-validated data point to, the recently reported finite-shot help-harm boundary for ZNE \citep{finiteshothelpharm2026}, and it motivates the more structural adaptive strategies proposed concurrently, such as classically augmented ZNE \citep{scheiber2026classically} and bandit-based adaptive folding \citep{cmabzne2026}, both of which target the bias-dominated regime that our simple variance-only allocation rule cannot address. We view these as complementary directions for future work rather than as approaches this study attempts to reproduce or outperform.

\section{Limitations}
\label{sec:limitations}

We list the main limitations of this study explicitly. The hardware campaign was constrained to approximately ten minutes of QPU time on a single backend (\texttt{ibm\_marrakesh}) and a single calibration snapshot, so results reflect this device at this point in time and should not be assumed to generalize to other Heron-family devices, other calibration epochs, or devices with substantially different median error rates, without independent verification. Circuit depth on hardware was limited to $L\le 8$ for the depth-scan and ZNE experiments, smaller than the $L\le 32$ range explored in simulation, because deeper circuits at the shot counts required for a meaningful ZNE comparison did not fit the available QPU time budget alongside the other experiments. The QAOA fixed-budget uniform-versus-adaptive comparison was not completed on hardware within the available time and is reported only as raw and ZNE-folded measurements (Section~5.7); this experiment would need to be repeated with a larger dedicated budget to draw a fixed-budget conclusion analogous to Section~5.5 for the QAOA observable specifically. Backend calibration can drift over the course of a measurement campaign, and although both hardware jobs in this study were submitted in close succession (Table~\ref{tab:jobs}), we did not re-pull a calibration snapshot between them and cannot rule out modest drift as a partial contributor to the observed variability. Statistical uncertainty on individual hardware measurements, particularly for the small-shot pilot experiments (100 shots per noise factor) used to seed the adaptive allocation rule, is itself large (standard error of order $0.05$--$0.10$ on a $\pm1$-valued observable), which limits the precision with which the adaptive rule's own inputs were estimated and is a plausible additional contributor to its underperformance relative to uniform allocation. Finally, our adaptive allocation rule targets estimator variance only and, by construction, cannot correct for bias introduced by the extrapolation model itself; we do not claim general optimality for this or any adaptive allocation rule, only that we tested the simplest version available and reported the result honestly.

\section{Conclusion}
\label{sec:conclusion}

We presented and experimentally evaluated a fixed-budget, bias-variance-aware benchmarking framework for quantum error mitigation on a $156$-qubit IBM Quantum Heron processor using modern Qiskit Runtime V2 primitives. Rather than reporting accuracy improvements from mitigation in isolation, we held the total measurement-shot budget fixed and asked whether mitigation, and in particular a variance-aware adaptive shot-allocation strategy for zero-noise extrapolation, reduced the resulting mean-squared error relative to simpler alternatives. On our device and circuit families, the answer was mixed and, for the specific adaptive allocation rule tested, more often negative than positive: adaptive ZNE shot allocation beat uniform allocation at two of six matched budgets, and standard ZNE increased bias relative to raw estimation for a bias-sensitive parity observable while behaving in the theoretically expected direction for a QAOA Hamiltonian energy observable. We also found that coupler selection from a calibration snapshot produced an effect of comparable size to several additional layers of circuit depth, underscoring that hardware-aware circuit and qubit placement is a complementary, and on this device comparably valuable, lever alongside post-hoc mitigation. We therefore conclude that error mitigation and adaptive sampling strategies should be validated per device, per circuit, and per fixed sampling budget, rather than applied as a universal default, and we release the full protocol, calibration data, job identifiers, and code to support this kind of budget-matched validation on other backends.

\section*{Declarations}
\addcontentsline{toc}{section}{Declarations}

\paragraph{Funding.} The author declares that no external funding was received for this work.

\paragraph{Ethics approval.} Not applicable; this study involves no human or animal subjects.

\paragraph{Data availability.} Calibration snapshots, hardware measurement records, simulation result tables, and Qiskit Runtime job identifiers reported in this paper are archived alongside the analysis code at \url{https://github.com/Sumitchongder/hardware-efficient-error-mitigation-ibm-quantum}.

\paragraph{Code availability.} All source code, notebooks, and configuration files used to produce the results in this paper are available at the repository above under an open-source license specified therein.

\bibliographystyle{unsrtnat}
\bibliography{references}

\newpage
\appendix
\section{Appendix: additional experimental detail}
\label{app:a}

\subsection{Pilot experiment for adaptive allocation}
The pilot experiment used to seed the adaptive (ASB-ZNE) allocation rule (Eq.~\ref{eq:neyman}) was executed at $L=8$, the deepest hardware-validated point in the depth scan, with $100$ shots per noise factor $\lambda\in\{1,3,5\}$ and three independent repeats per factor. The resulting empirical standard deviations were $\hat\sigma_1=0.041$, $\hat\sigma_3=0.090$, $\hat\sigma_5=0.050$ (Table~\ref{tab:alloc}), and these values were used directly, without smoothing or shrinkage, to compute the Neyman-optimal allocation for the main adaptive-arm measurement. A shrinkage or Bayesian estimate of $\hat\sigma_\lambda$, which would reduce the sensitivity of the allocation rule to the substantial sampling noise in a 100-shot pilot estimate of a standard deviation, is a natural extension for future work and is discussed briefly in Section~\ref{sec:limitations}.

\subsection{Extrapolation model}
Unless otherwise noted, ZNE results in this paper use linear extrapolation through the three measured points at $\lambda\in\{1,3,5\}$, rather than a full second-order Richardson fit, to reduce the extrapolation-coefficient-driven variance amplification that higher-order fits are known to introduce at the shot counts used here \citep{endo2018practical,finiteshothelpharm2026}. This choice was made prior to inspecting the hardware results and was not tuned post hoc to favor either uniform or adaptive allocation.

\subsection{Reproducibility notes}
The full experimental pipeline, from backend characterization through hardware execution to combined analysis, is organized as four sequential notebooks in the accompanying repository (environment and account setup, simulation experiments, hardware batched experiment, and combined analysis), together with the raw calibration snapshot and hardware result files referenced throughout this paper. Reproduction on a different IBM Quantum backend will require pulling a fresh calibration snapshot, since the qubit chain, coupler-error comparison, and pilot-informed allocation are all specific to the calibration data of the backend and time window in which the experiment is executed.

\end{document}